\documentclass[a4paper,fleqn,usenatbib]{mnras}

\usepackage{newtxtext,newtxmath}
\usepackage[T1]{fontenc}
\usepackage{ae,aecompl}

\usepackage{graphicx}	% Including figure files
\usepackage{amsmath}	% Advanced maths commands
\usepackage{amssymb}	% Extra maths symbols

\title[Do bulges stop stars forming?]{Do bulges stop stars forming?}

\author[S.A. Eales et al.]{Stephen Eales$^{1,2}$\thanks{E-mail: steve.a.eales@gmail.com},
Oliver Eales$^3$ and Pieter de Vis$^1$\\
$^{1}$ School of Physics and Astronomy, Cardiff University, The Parade, Cardiff CF24 3AA, UK\\
$^{2}$ Institute of Astronomy, University of Cambridge, Madingley Road, Cambridge, CB3 0HA, UK\\
$^{3}$ Department of Infectious Disease Epidemiology, School of Public Health, Faculty of Medicine,\\ 
Imperial College London, Medical School Building, St Mary's Campus, Norfolk Place,
London W2 1PG, UK\\ 
}

\date{Accepted XXX. Received YYY; in original form ZZZ}

\pubyear{2015}

\begin{document}
\label{firstpage}
\pagerange{\pageref{firstpage}--\pageref{lastpage}}
\maketitle


\begin{abstract}
In this paper, we use the {\it Herschel} Reference Survey to make a direct test of the
hypothesis that the growth of a stellar bulge leads to a reduction in the star-formation
efficiency of a galaxy (or conversely a growth in the gas-depletion timescale) as a result of
the stabilisation of the gaseous disk by the gravitational field of the bulge.
We find a strong correlation between star-formation efficiency and specific star-formation
rate in galaxies without prominent bulges and in galaxies of the same morphological type,
showing that there must be some other process besides the growth of a bulge that reduces the star-formation
efficiency in galaxies.
However, we also find that galaxies with more prominent bulges
(Hubble types E to Sab) do have significantly lower star-formation efficiencies than
galaxies with later morphological types, which is at least consistent with
the hypothesis that the growth of a bulge leads to the reduction in the star-formation
efficiency.
The answer to the question in the title is therefore, yes and no:
bulges may reduce the star-formation efficiency in galaxies but there must also be
some other process at work. We also find that there is a significant but small
difference in the star-formation efficiencies of galaxies with and without bars,
in the sense that galaxies with bars have slightly higher star-formation efficiencies.
\end{abstract}


%%%%%%%%%%%%%%%%%%%%%%%%%%%%%%%%%%%%%%%%%%%%%%%%%%

%%%%%%%%%%%%%%%%% BODY OF PAPER %%%%%%%%%%%%%%%%%%

\section{Introduction}

When investigating galaxy evolution astronomers have the big advantage over real historians or archaeologists that
they can see galaxies in the past. This sometimes seems a rather tantalizing advantage because we can only see a galaxy
at one moment and cannot watch the evolution of an individual galaxy. In practice, we are forced to
adopt a statistical approach.
The first step in this approach is
the uncontroversial, although challenging, one of building up, through observational programmes, a statistical description
of the galaxy population at each cosmic epoch. The second step, however, is subject to - and actually requires -
theoretical bias
because it involves {\it inferring} the physical connections between the galaxy populations at the different epochs.

For example, suppose that there is a class of galaxies at $\rm z=1$ 
that has very similar properties to a class of
galaxies at $\rm z=0$. To give a concrete example, let us imagine
that the low-redshift and high-redshift classes both contain
galaxies 
with an exponential stellar disk.
Does this similarity imply that the galaxies in these classes haven't evolved very much? The 
answer here is obviously no, because it is perfectly possible that a galaxy with an exponential disk
at $\rm z=1$ might now have a very different stellar structure, and a galaxy with an exponential disk
at $\rm z=0$ might have had a very different
structure at a $\rm z=1$. Inferring the physical connections is
not only challenging but is actually impossible without some prior theoretical assumption about
how galaxies evolve. 
On
top of this rather abstract problem, there is also the very practical problem that many of the basic
physical properties of
galaxies are remarkably difficult to measure. For example, there are not many properties more
basic than the star-formation rate (SFR) in a galaxy, but
there are at least 12
different methods that have been proposed for measuring the SFR, none of which is
entirely satisfactory (Kennicutt and Evans 2012; Davies et al. 2016).

In view of the difficulty of measuring such basic physical properties, much of the work in the field
over the last few decades has been based on statistical descriptions using simple observational
parameters. The classic example are the diagrams of optical colour versus absolute magnitude plotted
for galaxies found in large optical surveys. The galaxies found in these surveys fall in two main areas
of these diagrams: a narrow band occupied by galaxies with red colours called 
the `red sequence' and a more extended region of galaxies with blue colours, the
`blue cloud', separated by
the `green valley' where there are fewer
galaxies (Baum 1959; Strateva et al. 2001; Bell et al. 2003; Baldry et al. 2012). The red sequence
is mostly but not completely (Cortese et al. 2012 and references
therein) composed of galaxies with morphologies in the early-type class
(ellipticals and S0s) whereas the blue cloud is mostly composed of late-type galaxies (irregulars or
disk-dominated galaxies). This simple observational picture has given rise to the natural 
conclusion that there are two distinct classes of galaxy: star-forming galaxies in the
blue cloud and galaxies in which there is now very little star formation - 
variously called `red and dead', `quiescent'
or `quenched'
galaxies - forming the red sequence. 
A consequence of this dichotomy is that there must be some
physical process that converts a star-forming galaxy in the blue cloud
into a galaxy with very little star formation on the red sequence, and this process
must have quenched the star-formation quickly (at least quickly
relative to the age of the Universe) to explain the relative dearth of galaxies in the green valley,
although there is no consensus
about the
identity of this quenching process (Eales et al. 2018a and references therein).

In a series of three papers, we have argued for an alternative model
of galaxy evolution. A hint that the picture above is not correct is that the galaxies
found in a submillimetre survey do not lie on a colour-absolute-magnitude diagram 
in the classic blue cloud and red sequence but 
instead have almost the opposite distribution, forming
a `green mountain' (Eales et al. 2018b).
This result suggests, at least, that the current picture of galaxy evolution may
be an accident of the fact that our knowledge of the galaxy population has hitherto
been almost entirely based on the galaxies found in optical surveys.

In the first paper in the series, we used the {\it Herschel} Reference Survey
(HRS)
to show that
on a diagram
of specific star-formation rate (star-formation
rate divided by stellar mass; sSFR) versus stellar mass - meaningful physical properties
of galaxies rather than simple colours and absolute magnitudes - galaxies lie on a single
curved Galaxy Sequence (GS) rather than the two separate distributions of star-forming
and red-and-dead galaxies suggested by the current paradigm (Eales et al. 2017). 
The morphologies of galaxies change gradually along the GS
rather than there being an obvious break between late-type and early-type galaxies (Eales
et al. 2017).
In another paper we showed that a single continuous GS in this plot of sSFR versus
stellar mass naturally leads to the distinctive distributions in
the colour-absolute-magnitude space as a result of the different selection effects
operating in the optical and submillimetre wavebands (Eales et al. 2018b). 
This picture of a single GS is also consistent with recent
studies 
that show the kinematic properties of
galaxies appear to change gradually along the Hubble sequence rather
than there being an abrupt change at the early-type/late-type boundary
(Cappellari et al. 2013;
Cortese et al. 2016). 

The single GS, and the gradual variation in galaxy properties along it,
does not require a rapid quenching process, in contrast to the standard
picture of a `star-forming main sequence' and separate region
of red-and-dead galaxies.
We showed that the curvature of the GS and the rapid low-redshift
evolution 
of the galaxy population seen in radio and submillimetre surveys can
be produced by a weak quenching process in which the gas supply to galaxies is
turned off, with the curvature of the GS and the 
evolution being
produced as the galaxies gradually use 
up their remaining gas (Eales et al. 2018a).

\begin{figure}
\includegraphics[width=70mm]{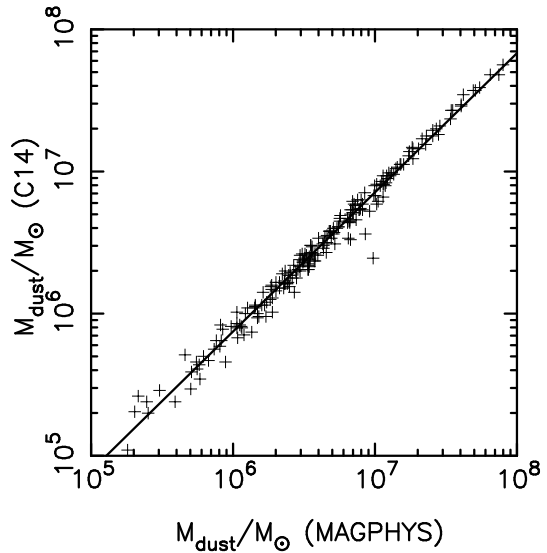}
  \caption{Dust mass estimated by MAGPHYS versus dust mass
estimated by Cortese et al. (2014; C14) using a single-temperature modified blackbody
for the 204 HRS galaxies with detections in all five {\it Herschel}
bands. The C14 dust masses
have been corrected to the Hubble constant used in the
paper and the
dust-mass opacity coefficient used in this MAGPHYS.
The line is the best-fit straight line
and has the form $\rm log_{10} M_{d, C14} = 0.979 log_{10} M_{d,MAGPHYS} + 1.91 \times 10^{-5}$.
The root mean squared variation around the line in the y-direction is 0.067.
}
\end{figure}

Another galaxy property that also seems
to vary gradually along the GS
is star-formation efficiency (star-formation rate divided by gas mass; SFE)
or its reciprocal, the depletion time (gas mass divided
by star-formation rate). A large number of studies, both of galaxies
at low redshift (Saintonge et al. 2011, 2012, 2017; Eales et al. 2018a) and 
high redshift (Scoville et al.
2017; Tacconi et el. 2018) have found that SFE and sSFR are strongly correlated,
with galaxies with lower values of sSFR also having lower values of SFE.
There is no consensus about the physical cause of this correlation but one appealing possibility
is that it is connected to the morphological change along the GS. Martig et al. (2009)
have suggested that the
star-formation efficiency will be less in a galaxy with a large bulge-to-disk ratio, since the
gas in the galaxy's disk will be stabilised against collapse into stars by
the
gravitational field of the bulge (see Figure 3 of that paper).
A corollary of that work is that the
decrease in SFE at low values of sSFR follows 
from the fact that
galaxies with low values of SSFR also tend to have more prominent bulges.

There is some indirect observational evidence for this hypothesis.
Saintonge et al. (2012) showed that galaxies 
in the GASS and COLD GASS surveys that have low values
of star-formation efficiency also have high values of the SDSS concentration index
and central surface density, both of which are characteristic of galaxies 
with high bulge-to-disk ratios.
Genzel et al. (2014) have used high-resolution observations of H$\alpha$ to
show that in the central regions of high-redshift galaxies the star-formation
rate is often lower than in the surrounding regions, the surface-density is high, and the
gas is stable against gravitational collapse
 - all of which is consistent
with the hypothesis that the growth of a central bulge in these galaxies has stabilised
the gas and reduced the star-formation efficiency.

This paper describes a  more direct attempt to test this hypothesis. The studies of the relationship
between SFE and sSFR cited above all used samples of galaxies that are too distant
for the morphologies of the galaxies to be classified. The HRS, however, contains galaxies
that are close enough for the morphologies of almost all of the galaxies to be known
(Boselli et al. 2010). By using the HRS galaxies, we can make a simple test of
the hypothesis that the reduction in the star-formation efficiency in
a galaxy is entirely the result of the growth of a bulge.

We assume a Hubble constant of 67.3 $\rm km\ s^{-1}\ Mpc^{-1}$
and the other {\it Planck} cosmological parameters (Planck Collaboration
2014). 

\section{Methods}

\subsection{The Herschel Reference Survey}

The sample we used in this paper is the {\it Herschel} Reference Survey (HRS; Boselli et al. 2010).
The HRS consists of 323 galaxies with
distances between 15 and 25 Mpc and with a near-infrared K-band limit of $\rm K <8.7$
for early-type galaxies (E, S0 and S0a - henceforth ETGs) and $\rm K <12$ for late-type galaxies (Sa-Sd-Im-BCD -
henceforth LTGs).
The sample was designed to select most of the galaxies
above stellar mass limits (different for ETGs and LTGs) in a given
volume of space
for an observing programme with the {\it Herschel} Space Observatory(Pilbratt et al. 2010).
The different K-band limits for ETGs and LTGs were chosen to avoid the sample being dominated by 
low-mass ETGs, which would have been hard to detect with {\it Herschel}. 
Eales et al. (2017) estimate that the HRS contains all LTGs in
the HRS volume with stellar masses above $\rm \simeq 8 \times 10^8\ M_{\odot}$ and
all ETGs above $\rm \simeq 2 \times 10^{10}\ M_{\odot}$.
Despite the high mass limit for ETGs, Eales et al. also show that $\simeq$90\% of the stellar
mass in this volume in ETGs with stellar masses $\rm > 10^8\ M_{\odot}$ is contained
in the galaxies included in the HRS. The reason for this low value is the turnover
in the stellar mass function for ETGs (Baldry et al. 2012), which means that
although ETGs with low stellar masses are missing from the HRS, they actually
contain only a small percentage of the total stellar mass in the ETG class.
The HRS therefore contains most of the stellar mass in a given volume of space, and
the size of this volume means that it is a fair sample of what has been produced by
12 billion years of galaxy evolution, with the only caveat being that the volume contains
the Virgo cluster and so cluster galaxies may be over-represented.

\begin{figure*}
\includegraphics[width=140mm]{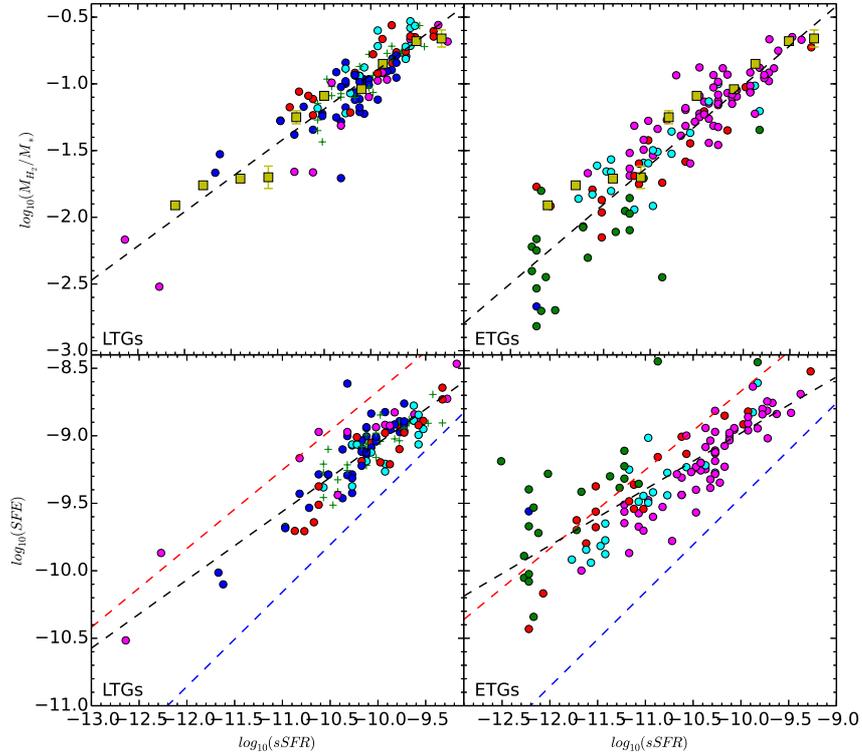}
  \caption{The two top panels show the ratio of gas mass to stellar mass plotted
against specific star-formation rate (sSFR), the two bottom panels star-formation
efficiecy (SFE) plotted against sSFR. The gas masses
have been calculated using equation 1.
The two left-hand panels are for LTGs, which are here defined
as Sc galaxies and morphological
types that are later along the Hubble sequence. The two right-hand panels show ETGs, here defined
as morphological types
on the Hubble sequence between, and including, Sbc and E. In the left-hand
panels, the colour key is as follows: blue - Sc; green - Scd; red - Sd; cyan - Sm and Sdm;
magenta - I, I0 or Im. In the right-hand panels, the colour key is as follows:
blue - E and ES0; green - S0 and S0a; red - Sa; cyan - Sab; magenta - Sb and Sbc.
The beige squares in the top panels show the average value of gas mass divided by stellar mass 
for the galaxies in the xCOLD GASS survey (Table 6 of Saintonge et al. 2017).
The dashed black lines in the panels show the straight lines
that best fit the data, the equations for which are given in Table 1.
The dashed red lines shows the relationship found at zero redshift between
SFE and sSFR by Scoville et al. (2017). The dashed blue lines
show the approximate relationship found between SFE and sSFR
for the xCOLD GASS survey by Saintonge et al. (2017).
}
\end{figure*}

For the purpose of this paper, we need to able to estimate the
mass of the interstellar medium (ISM) in each HRS galaxy.
Although many of the HRS galaxies have been observed in the CO 1-0
and 21-cm lines, the standard methods of estimating the mass of the
molecular and atomic phases of the ISM, only about 70\% of 
the HRS galaxies have been observed in the CO 1-0 line
(Boselli et al. 2014).

Fortunately, many authors have argued that estimating the mass of the ISM in
a galaxy from the continuum dust emission is actually better than the
standard method
(e.g. Eales et al. 2012; Magdis et al. 2012; Scoville et al. 2012). 
The main advantages are: (1) the uncertainty in the X factor used to connect the
intensity of the optically-thick CO 1-0 line to the column density of molecular hydrogen;
(2) the evidence from {\it Fermi} (Abdo et al. 2010), {\it Herschel} (Pineda
et al. 2013) and {\it Planck} (Planck Collaboration 2011)
that one third of the molecular gas in our galaxy is not emitting CO; (3) the evidence
for a much simpler relationship between metallity and the dust-to-gas ratio
than for the CO X factor (Sandstrom et al. 2013; R\'emy-Ruyer
et al. 2014).
For the HRS we are much better placed to use this alternative method
because
all the HRS galaxies were observed with {\it Herschel} in five far-infrared
or submillimetre bands 
(Ciesla et al. 2012; Smith et al. 2012; Cortese
et al. 2014). Of the 323 HRS galaxies, 284 (88\%) were detected in at least one band.
In this paper, we adopt this alternative method of estimating the mass of the
ISM.

\begin{figure*}
\includegraphics[width=140mm]{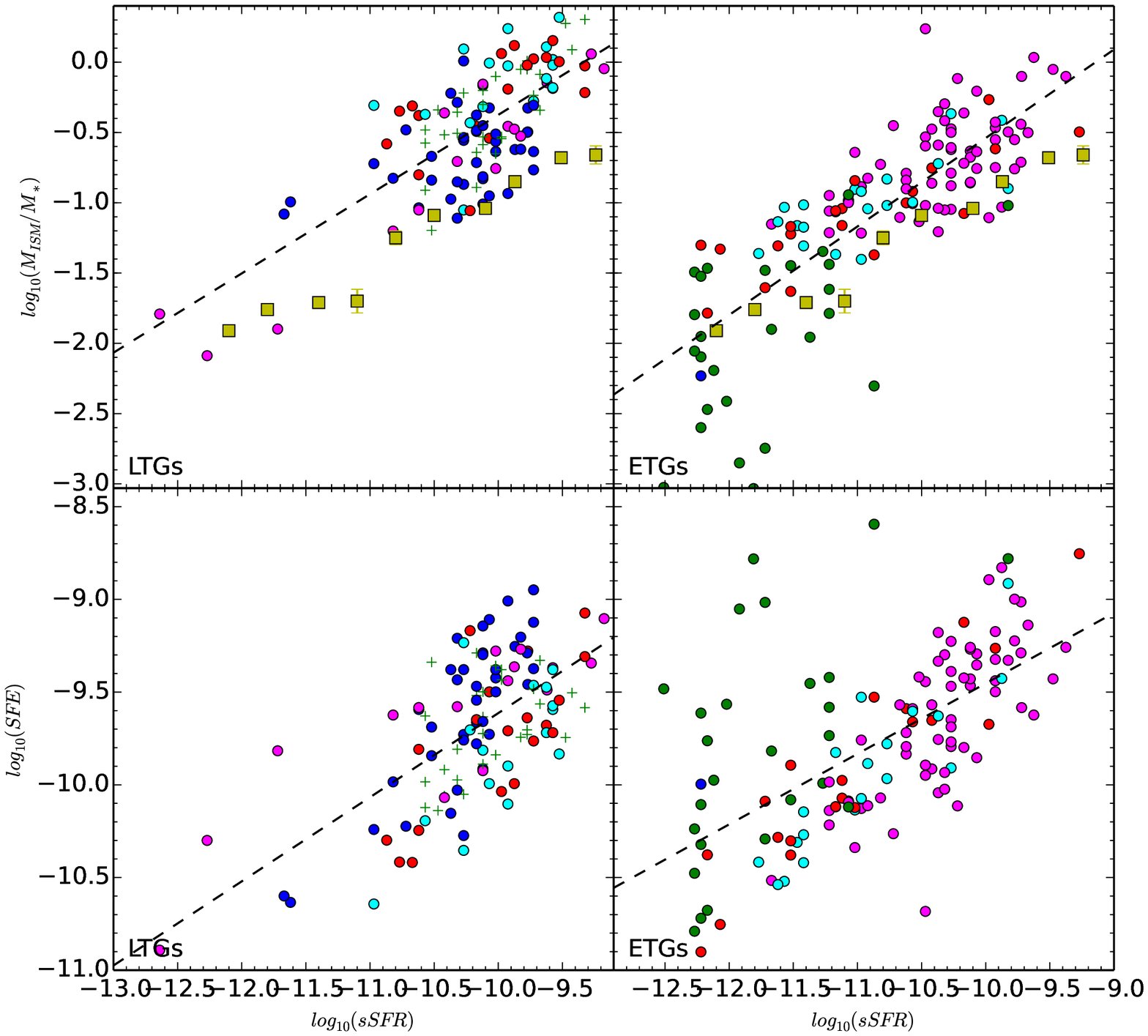}
  \caption{The same as Figure 2 except that the gas masses have been estimated
using equation 2.
}
\end{figure*}
\subsection{Star Formation Rates, Stellar Masses and Dust Masses}

The HRS galaxies have high-quality photometry in 21 photometric bands from
the $UV$ to the submillimetre (Eales et al. 2017 and references therein), making
it ideal for the application of a galaxy modelling program such as
MAGPHYS (Da Cunha et al. 2008). De Vis et al. (2016) applied MAGPHYS to
the HRS galaxies and we use their estimates of star-formation rates, stellar masses and
dust masses in this paper. We refer the reader to that paper for a
detailed description of the application of MAGPHYS to the HRS.

\begin{figure}
\includegraphics[width=90mm]{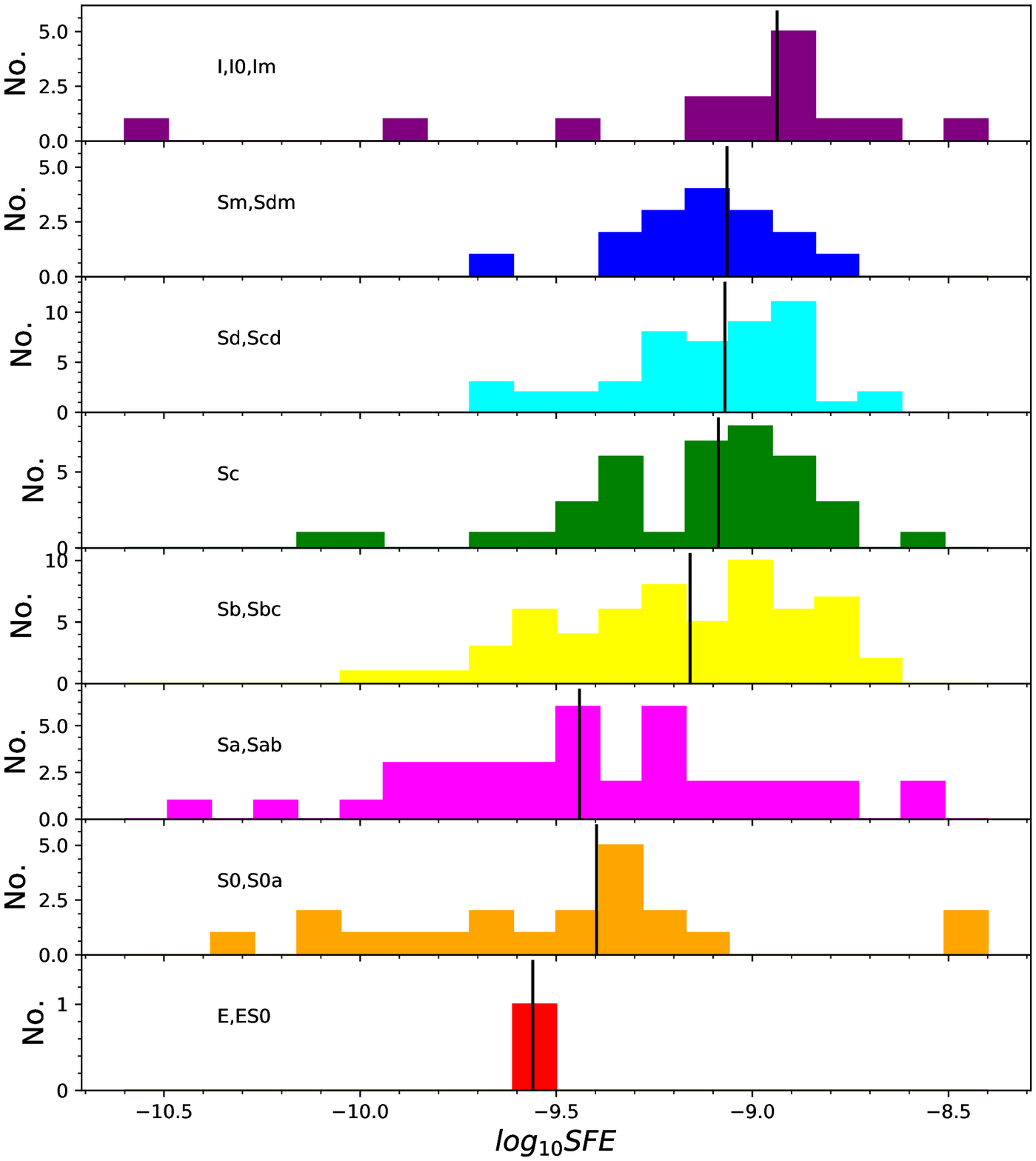}
  \caption{Histograms of star-formation efficiency for different morphological classes,
which are given by the label in each panel. The star-formation efficiency has been calculated
using the masses of molecular gas estimated using equation 1. The vertical line is
the median star-formation efficiency for each morphological class. 
}
\end{figure}

\begin{figure}
\includegraphics[width=90mm]{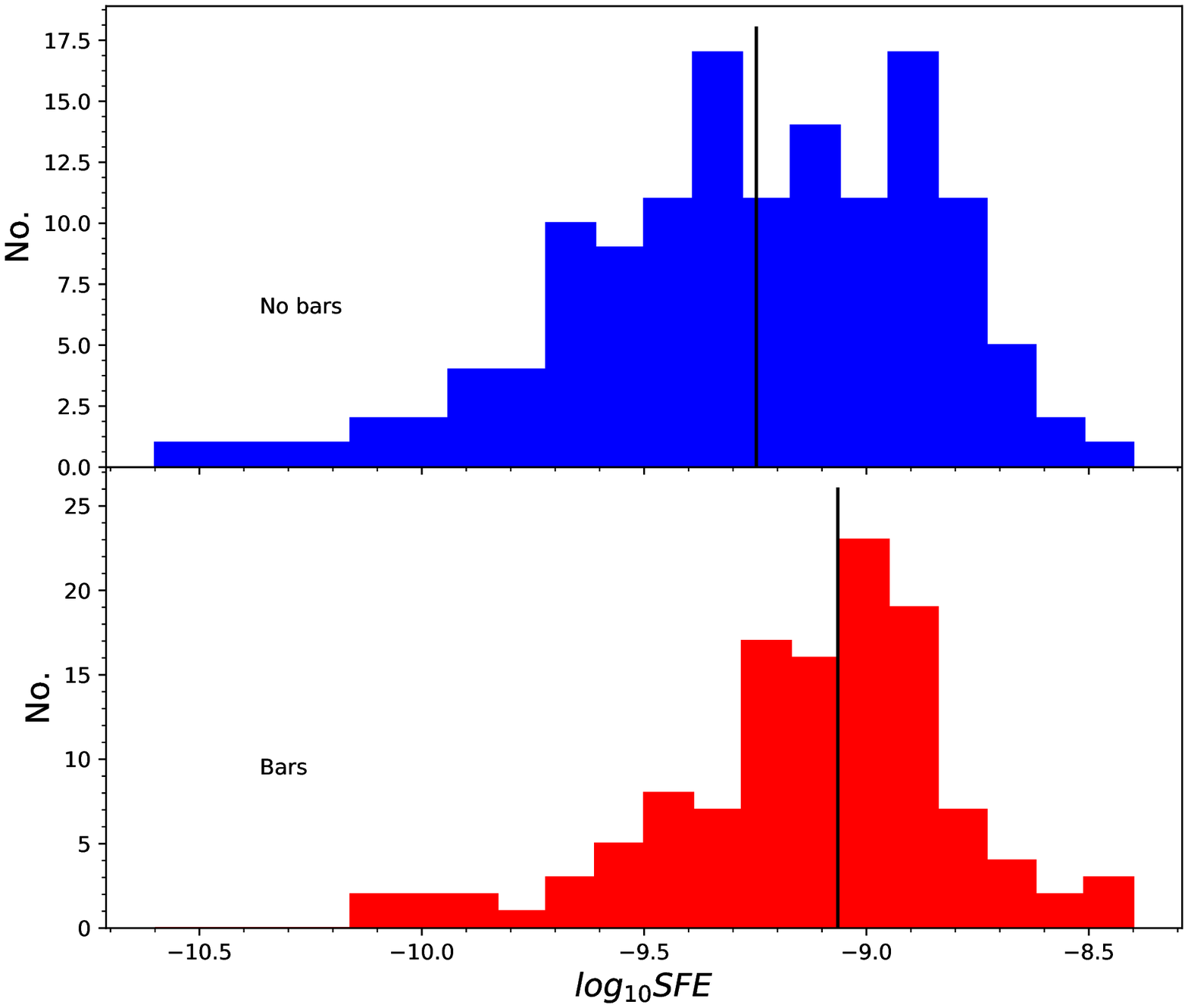}
  \caption{Histograms of star-formation efficiency for 
galaxies without bars (top) and with bars (bottom).
In this figure the barred galaxies include galaxies with
de Vaucouleurs morphological classes SB and SAB.
The star-formation efficiency has been calculated
using the masses of molecular gas estimated using equation 1. The vertical line is
the median star-formation efficiency for each class. 
}
\end{figure}

Very briefly, MAGPHYS generates 50,000 possible models of the spectral 
energy distribution of the unobscured stellar population, based on the stellar 
synthesis models of Bruzual and Charlot (2003), and 50,000 models of the dust emission from the ISM,
based on the ISM model of Charlot and Fall (2000).
The two sets of models are linked 
by balancing 
the energy absorbed by the dust in the optical and near-infrared with the energy emitted from the
dust in the mid-infrared 
to submm wavebands, which generates a large number of
possible spectral energy distributions (SEDs). These
are then fitted to the photometric measurements. 
For our project, an advantage of MAGPHYS is that it generates a probability distribution for
each galaxy property - in our case star-formation rate, stellar mass and dust mass - from the 
quality of the fits between the SEDs and the measurements.

We showed in Eales et al. (2017) that there
is good agreement between the stellar masses produced by
MAGPHYS and stellar masses estimated using a more empirical method.
We show here that there is also good agreement between the
MAGPHYS estimates of dust mass and empirical estimates.
Figure 1 shows a comparison of the MAGPHYS estimates 
with the estimates of dust mass made by Cortese et al. (2014)
for the 204 HRS galaxies that have detections in
all five {\it Herschel} bands, using fits of modified
blackbodies to the 
{\it Herschel} photometry.
We have corrected the latter estimates to
allow for the different Hubble constant used in this paper and the dust-mass opacity coefficient used
in MAGPHYS. The agreement is extremely good with the best-fit straight
line having a close to a unity slope ($\rm log_{10} M_{d, C14} = 0.979 log_{10} M_{d,MAGPHYS} + 1.91\times10^{-5}$)
and an root mean squared variation around the line of only 0.067. 

It is not possible to
test the MAGPHYS estimates of the star-formation rate in the
same way because there is no consensus about the best empirical
method of estimating the star-formation rate. 
A major advantage of MAGPHYS over most empirical methods
(Kennicutt and Evans 2012; Davies et al. 2016) is that it avoids the problem
of optically-based methods of missing the star formation that is
occurring deep in molecular clouds. One
obvious disadvantage is that it is based
on the 
assumption that galaxies are isotropic, but
Hayward and Smith (2015) have used simulated galaxies
to investigate the effect of viewing angle,
finding that when
a simulated disk galaxy is viewed from different directions
the variation in the MAGPHYS estimates of the star-formation rate 
is mostly within the $\rm \pm1\sigma$ error range estimated by
MAGPHYS.
Davies et al. (2016) also did an extensive comparison of the different methods that have 
been used for estimating star-formation rates, using 12 different methods to
estimate the star-formation rates for 4000 star-forming
galaxies. As their gold standard, 
they used the star-formation
rates estimated from the method of Popescu et al. (2011), which takes into account
the individual geometry of each galaxy.
They found that the slope of the relationship between star-formation rate and stellar mass
found by this method and by MAGPHYS were very similar. 

\subsection{Gas Masses}

We estimated the masses of gas in the HRS galaxies from the continuum dust emission
using two alternative approaches.

A number of groups have calibrated the
relationship between the luminosity of the continuum dust emission and the gas masses determined from
the standard CO 1-0 and 21-cm method (Eales et al. 2012; Scoville et al. 2014; Groves
et al. 2015; Janowiecki et al. 2018). An advantage of this method, rather than going through
the intermediate step of calculating the dust mass, is that it avoids the need to know
the value of the dust-mass opacity coefficient or the gas-to-dust ratio, neither of which
are known accurately (Clark et al. 2016). A disadvantage of the method is that it suffers
from the same limitations of the CO/21-cm method; if, for example, some fraction of the
molecular gas does not contain any CO, the dust method will also miss this gas.

Since Janowiecki et al. (2018) calibrated their relationship from 68 galaxies in the HRS,
we have used their relationship between the mass of molecular gas and the 500-$\mu$m luminosity:

\begin{equation}
log_{10}\left( {M_{H_2} \over M_{\odot}}\right) = 0.99\left(log_{10}{L_{500 \mu m} \over L_{\odot}} - 
9.0\right) + 8.41
\end{equation}

\smallskip

\noindent This method, of course, will produce an estimate of the mass of gas in the
molecular phase rather than the total mass of the ISM.

The second method we use was originally suggested by James et al. (2002)
but was recently updated by Clark et al. (2016). The method is based on the assumption,
for which there is a lot of observational support (Clark et al. 2016 and references therein),
that a constant fraction of the metals in a galaxy are in dust grains. If this
is the case, the mass of gas in a galaxy is given by:

\begin{equation}
M_g = {M_d \over \epsilon_d f_{Z_{\odot}} Z }
\end{equation}

\noindent where $\epsilon_d$ is the fraction of the mass of metals
in a galaxy that is incorporated in dust grains, $f_{Z_{\odot}}$ is the
metal mass fraction at solar metallicity, and $Z$ is the metallicity
of the ISM in a galaxy as a fraction of the solar value.
We have assumed that the value of $\epsilon_d$ is 0.5 (Clark et al.
2016) and a value for $f_{Z_{\odot}}$ of 0.0142. We have used the metallicities
measured for HRS galaxies by Hughes et al. (2013). For the HRS galaxies
without metallicity measurements, which are mostly ETGs, we estimated
the metallicity of the galaxy using the relationship between metallicity and
stellar mass derived by Hughes et al. (their equation 5). We have used the
estimates of dust mass produced by MAGPHYS.

A disadvantage of this method is that it relies on the value of the
dust-mass opacity coefficient used in MAGPHYS being correct. Errors in this, and
in any of the constants in equation 2,
will merely lead to all the gas masses being wrong by the same factor, and will
not lead to a spurious relationship between SFE and sSFR.

\section{Results}

Figure 2 contains four panels. The two upper panels show
the ratio of gas mass to stellar mass plotted against specific star-formation
rate (sSFR).
The two bottom panels show star-formation efficiency (SFE) plotted
against sSFR. In all panels, we have calculated the gas mass using equation 1.
The panels on the left of Figure 2
are for LTGs and the panels on the right are for ETGs. However, in this
figure we have redefined the two classes. In Figure 2 the ETGs are all those
galaxies with a prominent bulge, which we have defined as the morphological
types on the Hubble sequence between, and including, Sbc and E (121 galaxies). The LTGs are
all galaxies with morphological types Sc or later (118 galaxies). Figure 3 shows the
same four panels but this time with the gas mass calculated using
equation 2. In both figure the dashed black lines in each panel show the best fits of the
equation $log_{10}y = A log_{10} x + B$ to the data. The values of $A$ and $B$ are
given in Table 1, as are the values of the Spearman correlation coefficient for
each dataset.

\begin{table*}
\caption{Relationship between SFE and SSFR}
\begin{tabular}{ccccccc}
\hline
Relationship & Morphologies & ISM method & A & B & Spearman &  Reference \\
\hline
$M_{gas}/M_*$ versus sSFR & LTGs & CO & 0.51 & 4.20 & 0.858 & this paper \\
$M_{gas}/M_*$ versus sSFR & ETGs & CO & 0.59 & 4.93 & 0.840 & this paper \\
$M_{gas}/M_*$ versus sSFR & LTGs & MAGPHYS & 0.56 & 5.27 & 0.630 & this paper \\
$M_{gas}/M_*$ versus sSFR & ETGs & MAGPHYS & 0.63 & 5.76 & 0.825 & this paper \\
SFE versus sSFR & LTGs & CO & 0.51 & -4.00 & 0.822 & this paper \\
SFE versus sSFR & ETGs & CO & 0.42 & -4.82 & 0.840 & this paper \\
SFE versus sSFR & LTGs & MAGPHYS & 0.45 & -5.08 & 0.552 &  this paper \\
SFE versus sSFR & ETGs & MAGPHYS & 0.38 & -5.65 & 0.658 & this paper \\
\hline
SFE versus sSFR & All & ... & 0.58 & -2.84 & ... & Saintonge et al. (2017) \\
SFE versus sSFR & All & ... & 0.70 & -2.46 & ... & Scoville et al. (2017) \\
\hline
\end{tabular}
\end{table*}

Let us consider first the top panels of Figures 2 and 3, the plots of the ratio
of gas mass to stellar mass versus sSFR. 
The beige points in both panels show the average values of this ratio for the molecular
gas for galaxies
in the xCOLD GASS survey (Saintonge et al. 2017), which were not separated by
morphological type. In Figure 2 the agreement is good for both ETGs and LTGs.
In Figure 3 the points from xCOLD GASS are significantly lower than our points, although
we see the same trend between the mass ratio and sSFR. Apart from the possibility
that we are using the wrong values for the constants in
equation 2, the obvious explanation of the discrepancy is that in Figure 3 we are
using estimates of the entire mass of the ISM not just the molecular phase.

Now let us consider the bottom panels of both figures, the plots of SFE versus sSFR.
The blue dashed
line in Figure 2 shows the relationship found for the galaxies in xCOLD GASS (Saintonge et al. 2017) 
and the red dashed line
shows the relationship for zero redshift found by Scoville et al. (2017). We have not plotted
these relationships in Figure 3 because 
they only include the molecular phase of the ISM.
Strong relationships between SFE and sSFR are seen in our data in both Figures 2 and
3, for both
the ETGs and LTGs, with slopes similar to those found in the other studies, although
there are significant offsets (Table 1), which it seems likely
are caused by systematic
differences in the methods used to estimate gas mass
and star-formation rate.

At this point, it is important to consider whether all
the relationships might be spurious.
Both SFE and sSFR are ratios with star-formation rate as the numerator.
Errors in the star-formation rate might therefore
produce a spurious correlation between SFE and
sSFR, a possibility that, of course, applies to all previous studies of this
sort. 
Appendix A describes an investigation of this possibility.
We show that the errors do produce a correlation between SFE and sSFR
but with a much shallower slope than that seen in Figs 2
and 3.

We further investigated the relationships between the different
variables by using the Metropolis-Hastings algorithm
to fit the following two equations to the data (Appendix B).
Following Scoville et al. (2017), these relationships allow for the
possibility that stellar mass is an important variable.

\begin{equation}
M_{gas} = A\ M_{\odot} \times \left({ sSFR \over 10^{-10} } \right)^B \times 
\left({ M_* \over 10^{10} M_{\odot}}\right)^C
\end{equation}

\begin{equation}
SFR = A\ M_{\odot}\ year^{-1} \times
{M_{gas} \over 10^9\ M_{\odot}} \times \left({ sSFR \over 10^{-10} } \right)^B
\times 
\left({ M_* \over 10^{10} M_{\odot}}\right)^C
\end{equation}

\noindent The results are given in Table 2. The values derived for B in the second equation
are reassuringly similar to the values for the slope of the relationship
between SFE and sSFR given in Table 1.

\begin{table*}
\caption{Fits of Equations 3 and 4}
\begin{tabular}{ccccc}
\hline
Equation & Morphologies &  A & B & C \\
\hline
3 & LTGs & 9.04$\pm$0.01 & 0.50$\pm$0.01 & 0.96$\pm$0.01 \\
3 & ETGs & 8.97$\pm$0.01 & 0.57$\pm$0.01 & 0.95$\pm$0.01 \\
4 & LTGs & -0.038$\pm$0.004 & 0.507$\pm$0.006 & 0.029$\pm$0.006 \\
4 & ETGs & 0.067$\pm$0.003 & 0.486$\pm$0.004 & 0.031$\pm$0.005 \\
\hline
\end{tabular}
\end{table*}

\section{Discussion}

The correlations between star-formation efficiency (SFE) and specific
star-formation rate (sSFR) are much stronger in Figure 2 than Figure 3.
Since the gas masses used in Figure 2 only include molecular gas, whereas the
gas masses in Figure 3 incorporate all phases of the ISM in which dust is
found, the tighter correlation in Figure 2 is easily explained by
the fact that stars form out of molecular gas.

The most interesing result apparent in Figures 2 and 3 is that similar correlations are found
between SFE and
sSFR for both LTGs and ETGs. 
This is interesting because of the 
suggestion that a more prominent stellar bulge in a galaxy
stabilizes the gas in the surrounding disk, stopping it collapse
to form stars, thus reducing the value of the SFE
(Martig et al. 2009). 

Figures 2 and 3 contain two pieces of evidence
against this idea.
The first is
that very similar relationships are seen for LTGs and ETGs.
If bulges stop stars forming, we would not expect
to see a relationship for galaxies without prominent
bulges. The galaxies in our redefined LTG class with the most prominent
bulges are the Sc galaxies, but even for these
the ratio of the near-infrared luminosity of the bulge
to the luminosity of the galaxy itself, which
should be
similar to the bulge-to-total mass ratio, is only $\simeq$10\%
(Laurikainen et al. 2007).
The second piece of evidence is that even within a single morphological
class, all of which should have a similar bulge-to-disk ratio,
we see the same relationship between SFE and sSFR. 
The dark blue points in the LTG plots represent Sc galaxies and
the magenta points in the ETG plots represent Sb and Sbc
galaxies; within both classes there is
a strong correlation between SFE and sSFR.

Our first conclusion is therefore that the growth of a stellar bulge, and the consequent
stabilisation of the gas in the disk, is not enough to explain the reduction in the SFE seen in
galaxies that have low values of sSFR. There must be some other process at work.

Bluck et al. (2014) have looked at the importance of four different
parameters for predicting that a galaxy has a low value of sSFR: stellar mass,
halo mass, bulge-to-disk ratio and bulge mass. They find that the one that is
most important is bulge mass. Their conclusion that the bulge-to-disk ratio is
comparatively unimportant is 
consistent with
the lack of a strong correlation between Hubble type, which 
of course depends strongly on bulge-to-disk ratio (Laurikainen et al. 2007),
and sSFR in the
HRS
(Eales et al. 2017; see also Figs 2 and 3).
It also suggests that bulge-to-disk ratio cannot be the property that
explains the relationship between SFE and sSFR, since if it were we would expect to
see strong correlations between bulge-to-disk ratio and both SFE and sSFR.
Thus their work generally supports our conclusion that the disk-stablisation
idea is not enough to explain the relationship between SFE and sSFR.

Is there any evidence for the disk-stablisation hypothesis in the HRS? 
Both Figures 2 and 3 do suggest that the spread in SFE is larger for ETGs
than LTGs and it is the galaxies with the earliest morphological types have the
lowest values of SFE. Figure 4, in which we have plotted histograms of SFE
for galaxies in different morphological classes, shows this more clearly.
For this figure we have used the values of SFE calculated from the molecular
gas masses estimated using equation 1.
There is a gradual drop in SFE along the Hubble sequence from later to earlier morphological
types, but there seems
to be a particular large jump down in SFE between the fifth and sixth (from the top)
panels,
at the boundary between Sb and Sab. We have tested the significance of this
fall by treating all the galaxies in the top five histograms as one sample and all the
galaxies in the bottom three histograms as a second sample, and then applying the
Kolmogorov-Smirnov two-sample test. We find that the Kolmogorov-Smirnov statistic has a value
of 0.428 with a formal significance of $\rm 4 \times 10^{-8}$.

This agrees with the results of Saintonge et al. (2012). Although they did not have
morphological classifications for the galaxies in their sample, they showed
that galaxies with low values of SFE also tend to have high values for
the SDSS concentration index and central stellar surface-density, both of which
are generally higher for bulge-dominated galaxies. For completeness, since
Saintonge et al. also looked at the effect of bars on SFE, Figure 5 shows
histograms of SFE for galaxies with bars and galaxies without bars in the
HRS. In the figure we have lumped together galaxies with strong bars (de Vaucouleurs
type SB) and with weak bars (de Vaucouleurs type SAB), but it makes
little difference if we only put galaxies with strong bars in the
barred class. There is a significantly different (Kolmogorov-Smirnov statistic of
0.23, probability under null hyothesis of identical populations of 0.0018) but it
is not a large one. The median $\rm log_{10}(SFE)$ for the barred galaxies is -9.06 
and for the galaxies without bars it is -9.25. This also agrees well
with the results of Saintonge et al. (2012) who also found a significant but
small difference between the SFEs of galaxies with and without bars.

We emphasise that the difference in SFE seen in Figure 4 between the early-type and
late-type galaxies does not prove that it is the effect of the bulge in stabilising
the disk that is responsible for the reduction in SFE.
It is possible that the apparent relationship between SFE and morphological class
is the result of two separate relationships: a relationship between sSFR and morphological
class caused by some unknown process (perhaps bulges formed more readily at earlier cosmic
epochs, so galaxies in which most of the stars formed earlier, and thus today
have lower values of sSFR, have more prominent bulges) and a relationship between
SFE and sSFR caused by the other process for reducing SFE that we
have argued must exist.

The strong result of this paper is that
there must be some other process besides the disk-stabilising effect of
a bulge that must be reducing the SFEs in galaxies, although the
observations are consistent with the idea that the disk-stabilising process
is also occurring.
As to what this other process is, we don't know.
One speculative idea comes from considering the role of turbulence
in determining the SFE.
Increased turbulence seems to be associated
with an increase in the SFE of a galaxy, probably because
the increase in turbulence leads to an increase in the fraction
of gas that is in a very dense phase (Papadopoulos and Geach 2012).
Turbulence is injected into a galaxy by both the accretion of gas
and feedback from star formation
(Krumholz et al. 2018),
so the current velocity field in a galaxy
is the result of the history of both processes.
A plausible, but clearly very speculative, explanation
of the low values of SFE for galaxies with low values of sSFR is that the
main star-formation phase of these galaxies will have been
further back in the past, gving more time for the velocity field, stirred
up by the gas accretion and feddback during the star-formation phase,
to have quietened down.
A simple test of this hypothesis
would be to investigate whether any of the statistical measures of a
galaxy's velocity field depend on sSFR for galaxies in the universe today,
a test well within the capabilities of the Atacama Large Millimetre Array.

\section{Conclusions}

In this paper, we have used the 
{\it Herschel} Reference Survey to make a direct test of the
hypothesis that the growth of a stellar bulge leads to a reduction in the star-formation
efficiency (or conversely an increase in the gas-depletion timescale) as a result of
the stabilisation of the gaseous disk by the gravitational field of the bulge.
We find a strong correlation between star-formation efficiency and specific star-formation
rate in galaxies without prominent bulges and in galaxies of the same morphological type,
showing that there must be some other process besides the growth of a bulge that reduces the star-formation
efficiency in galaxies.

However, we also find that galaxies with more prominent bulges do tend to have lower
values of SFE, confirming the conclusion of Saintonge et al. (2012), who used a more indirect
method that did not use morphological classifications. Confirming another result of
Saintonge et al. (2012), we also find a significant but small difference between
the SFEs in galaxies with and without bars, in the sense that galaxies with bars have
slightly higher values of SFE.

Our strong conclusion is that there must be some other process beyond the disk-stabilising effect
of a bulge that is responsible for the reduction of the SFE in galaxies.
Our weaker conclusion is that the difference in SFE between different morphological
classes is at least consistent with the hypothesis that the growth of a bulge stabilises the
gas in the disk, thus reducing the SFE.

\section*{Acknowledgements}

We thank Asa Bluck for many useful discussions. We thank Asa Bluck
and Frederic Bournaud for comments on the manuscript that significantly
improved it.
SAE thanks Clare Hall in Cambridge for a Visiting Fellowship, and the college, the
Kavli Institute and the Institute of Astronomy for making his visit to Cambridge an
enjoyable and stimulating one.
He acknowledges funding  from  the  UK  
Science  and  Technology  Facilities  Council  consolidated  grant  ST/K000926/1.

%%%%%%%%%%%%%%%%%%%%%%%%%%%%%%%%%%%%%%%%%%%%%%%%%%



%\bibliographystyle{mnras}


%%%%%%%%%%%%%%%%%%%%%%%%%%%%%%%%%%%%%%%%%%%%%%%%%%

%%%%%%%%%%%%%%%%% APPENDICES %%%%%%%%%%%%%%%%%%%%%

\appendix

\section{Are relationships between SFE and SSFR caused by experimental errors}

Both star-formation efficiency (SFE) and specific star-formation rate (sSFR) 
are ratios with 
star-formation rate as the numerator in the ratio. Therefore errors in the
estimates of the star-formation rate could potentially generate spurious relationships
between SFE and sSFR.
We have tested this possibility using the following simple Monte-Carlo procedure.

In this test we used the molecular gas masses of Figure 2 rather than the
MAGPHYS gas masses of Figure 3. The simulation is based on the assumption
that all galaxies in a morphological class (LTGs or ETGs)
have the same value of SFE, which
we take as our mean estimate of the SFE for the galaxies in that class.
We then start with the MAGPHYS stellar mass and our estimate of the
gas mass of each galaxy and use the
assumed SFE to calculate a star-formation rate for that galaxy. In the absence of errors,
there would then be no correlation between SFE and sSFR. We generate an error
for each galaxy using the errors on $\rm log_{10} SFR$ given by MAGPHYS.
Figure A1 shows the result, which should be compared with Figure 2.

The figure shows that a correlation between SFE and sSFR has been generated
by the errors in the star-formation rate.
The dashed black lines in the figure show the straight lines that are
the best fit to the data, and
Table A1 lists the equations for these lines and the values of the Spearman
correlation coefficient.
The errors clearly produce a significant correlation but the slope of the
relationship between SFE and sSFR is much smaller than the
slope of the relationship seen in Figure 2.

It is possible, of course, that the errors in the star-formation rate
given by MAGPHYS are underestimates. We have investigated this possibility
by making the assumption that the true errors in our values of $\rm log_{10} SFR$
are 3 times greater than those given by MAGPHYS. Figure A2 shows the result.
This time there is a strong correlation between SFE and sSFR, as strong as
that seen in Figure 2. However, the relationship in the top two
panels between $\rm M_{gas}/M_*$ is now much weaker than those
in the top two panels of Figure 2. We therefore conclude that
that the relationships shown in Figure 2 are not the result of errors
in the star-formation rate generating correlated errors in SFE and
sSFR.

\begin{figure*}
\includegraphics[width=140mm]{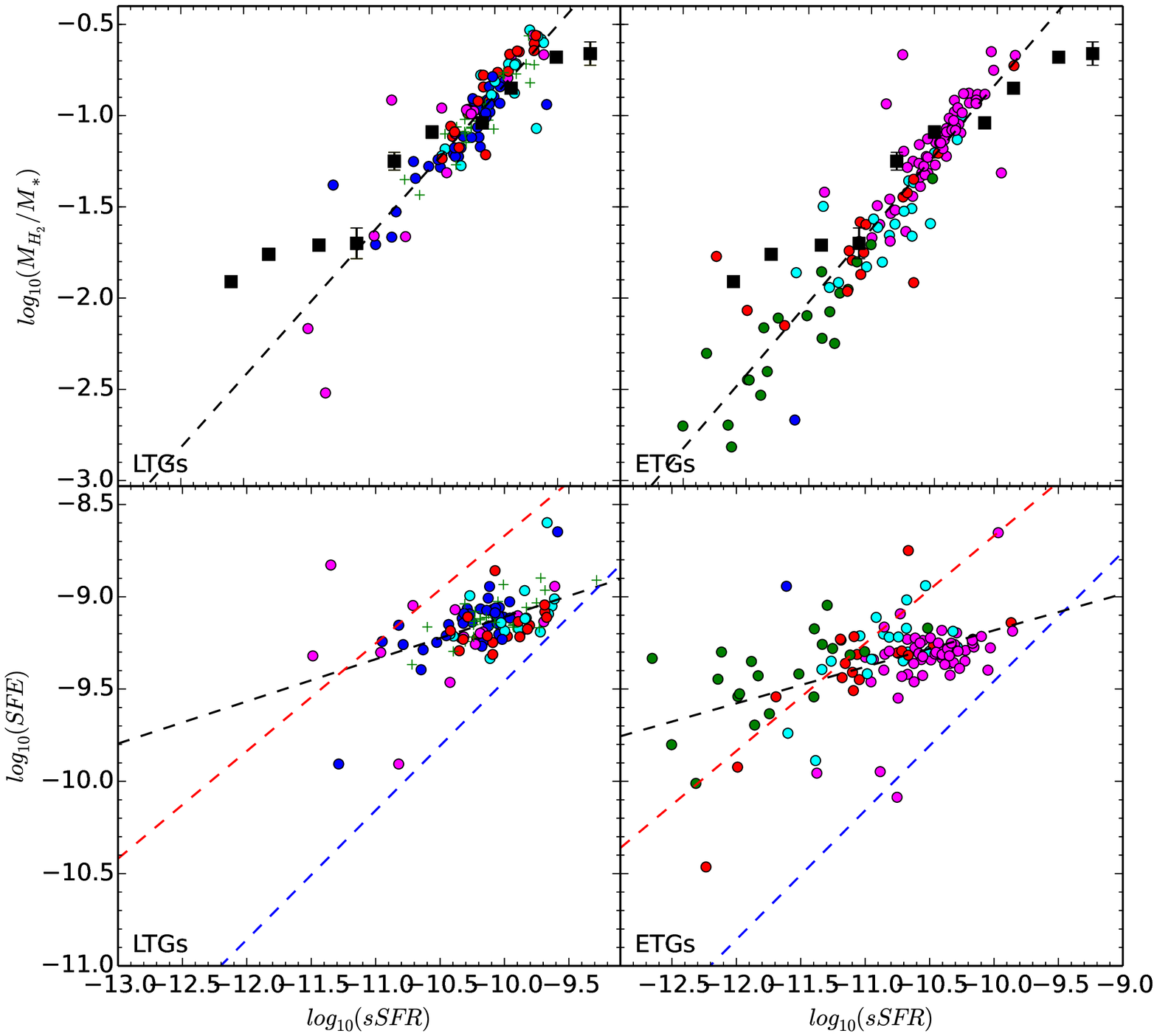}
  \caption{The same as Figure 2 except that the star-formation rates
have been generated using the Monte-Carlo method described in Appendix A,
which is based on the assumption that all galaxies in a morphological
class (LTGs or ETGs) have the same value of SFE. We have produced this
figure by making the assumption that the errors in the star-formation
rate are those given by MAGPHYS. The equations for the best-fit lines
(the dashed black lines) are given in Table A1.
}
\end{figure*}

\begin{figure*}
\includegraphics[width=140mm]{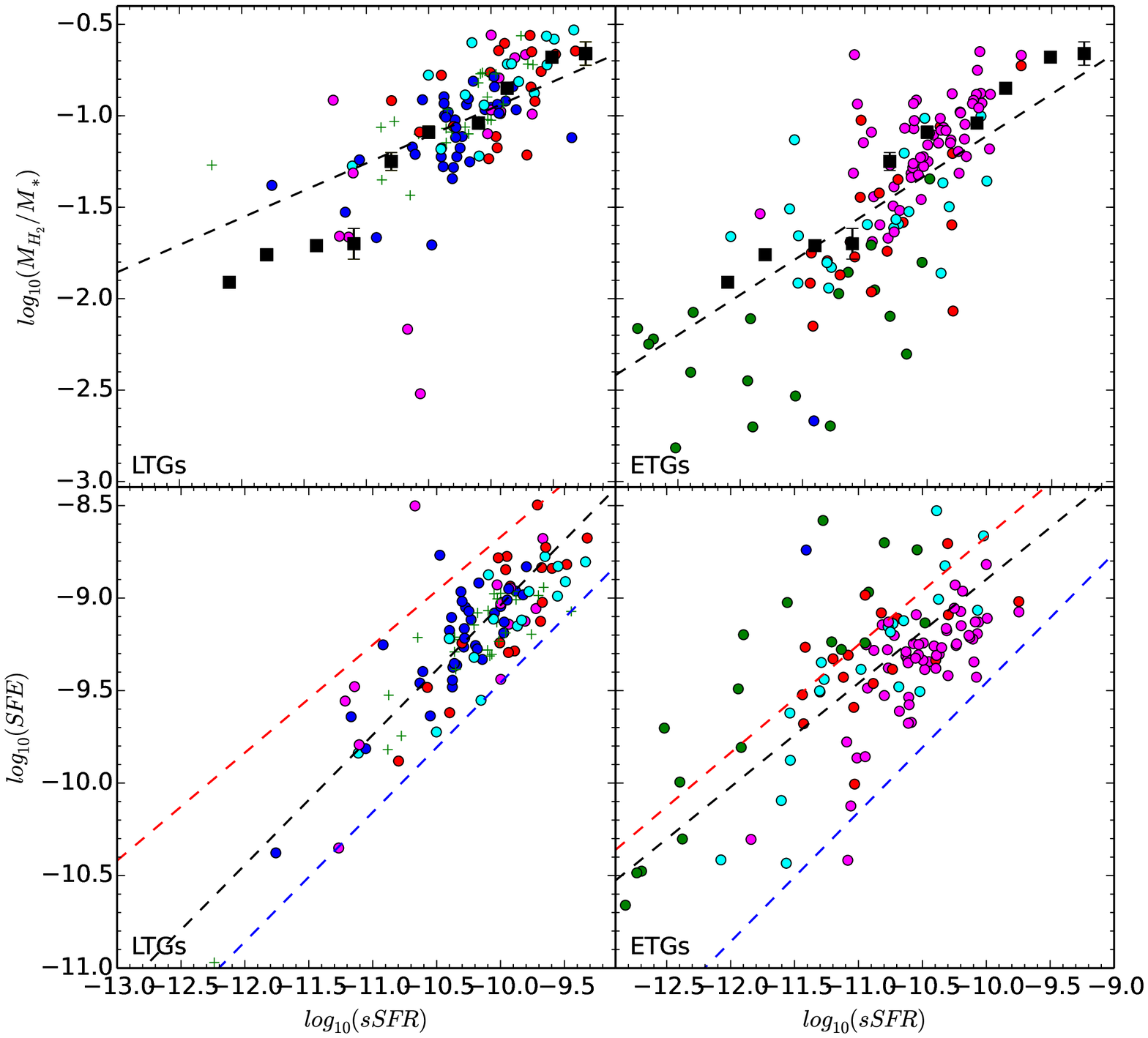}
  \caption{The same as Figure 2 except that the star-formation rates
have been generated using the Monte-Carlo method described in Appendix A,
which is based on the assumption that all galaxies in a morphological
class (LTGs or ETGs) have the same value of SFE. We have produced this
figure by making the assumption that the errors in $\rm log_{10}SFR$
are
three times those given by MAGPHYS.
The equations for the best-fit lines
(the dashed black lines) are given in Table A1.
}
\end{figure*}

\begin{table*}
\caption{Monte Carlo Results}
\begin{tabular}{cccccc}
\hline
Relationship & Morphologies & Errors & A & B & Spearman \\
\hline
$M_{gas}/M_*$ versus sSFR & LTGs & MAGPHYS & 0.77 & 6.82 & 0.884 \\
$M_{gas}/M_*$ versus sSFR & ETGs & MAGPHYS & 0.80 & 7.20 & 0.921 \\
SFE versus sSFR & LTGs & MAGPHYS & 0.23 & -6.82 & 0.470 \\
SFE versus sSFR & ETGs & MAGPHYS & 0.20 & -7.20 & 0.423 \\

$M_{gas}/M_*$ versus sSFR & LTGs & MAGPHYS$\times$3 & 0.37 & 2.74 & 0.616 \\
$M_{gas}/M_*$ versus sSFR & ETGs & MAGPHYS$\times$3 & 0.56 & 4.62 & 0.800 \\
SFE versus sSFR & LTGs & MAGPHYS$\times$3 & 0.63 & -2.74 & 0.689 \\
SFE versus sSFR & ETGs & MAGPHYS$\times$3 & 0.44 & -4.62 & 0.476 \\
\hline
\end{tabular}
\end{table*}

\section{Application of the Metropolis-Hasting technique
to the data}

We fitted equations 3 and 4 to the data 
by minimising chi-squared. We fitted equation 3 by minimising chi-squared
on the assumption that the errors in $log_{10} M_{gas}$ are 0.2.
We fitted equation 4 using the errors in $log_{10} SFR$ produced
by MAGPHYS. We sampled the posterior probabilty distribution of A, B and
C using the Metropolis-Hasting algorithm, with 500,000 points and a burn-in phase
of 2000 points. The probability distributions and covariance plots
for A, B and C are shown for late-type galaxies for the two equations
in Figures B1 and B2.

\begin{figure*}
\includegraphics[width=140mm]{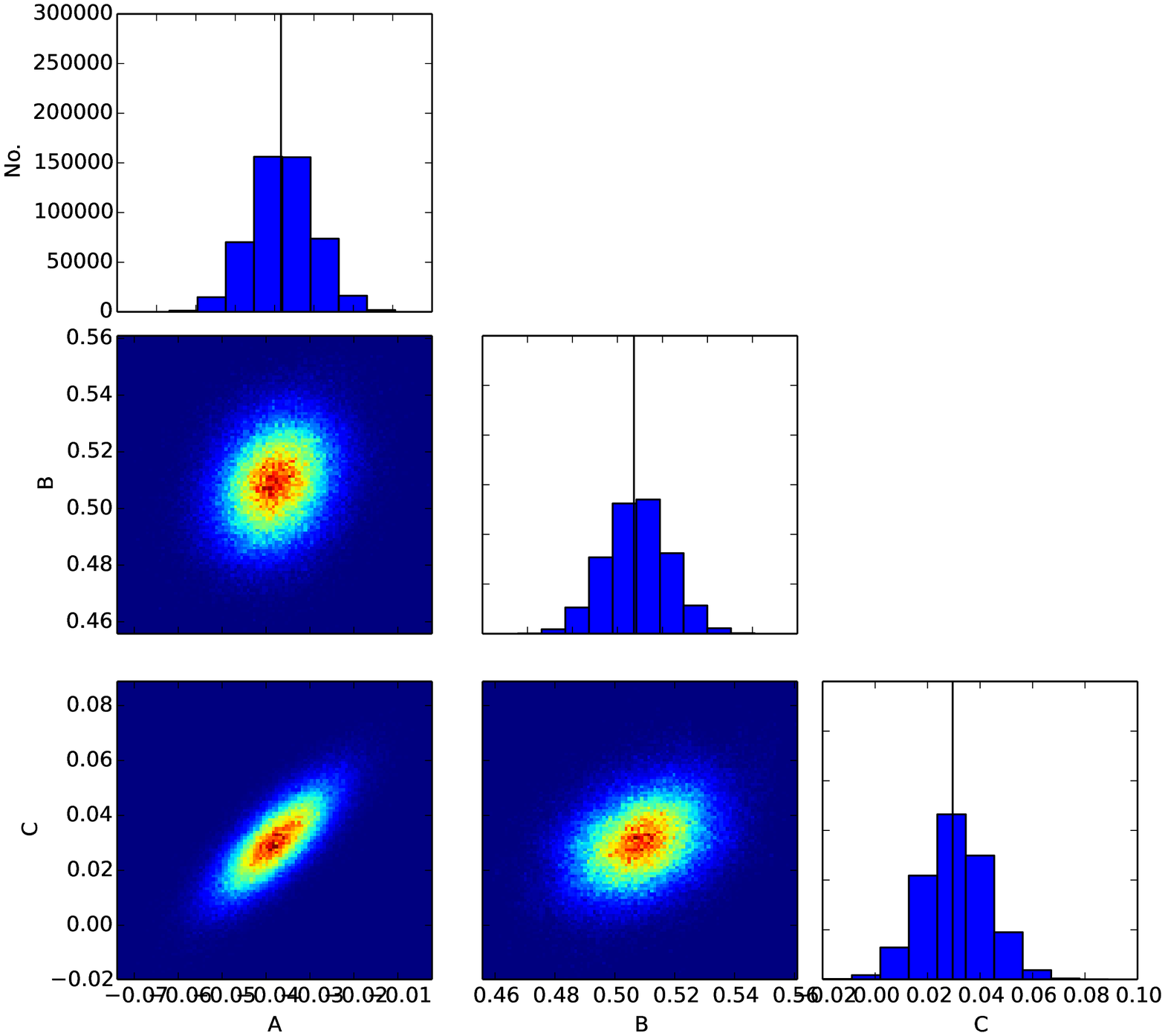}
  \caption{The results of fitting equation 3 to the data for the
late-type galaxies (as defined in Section 3).
The histograms show the probability distributions for the
three parameters, with the vertical lines showing the mean
values. The other panels show covariance plots.
}
\end{figure*}

\begin{figure*}
\includegraphics[width=140mm]{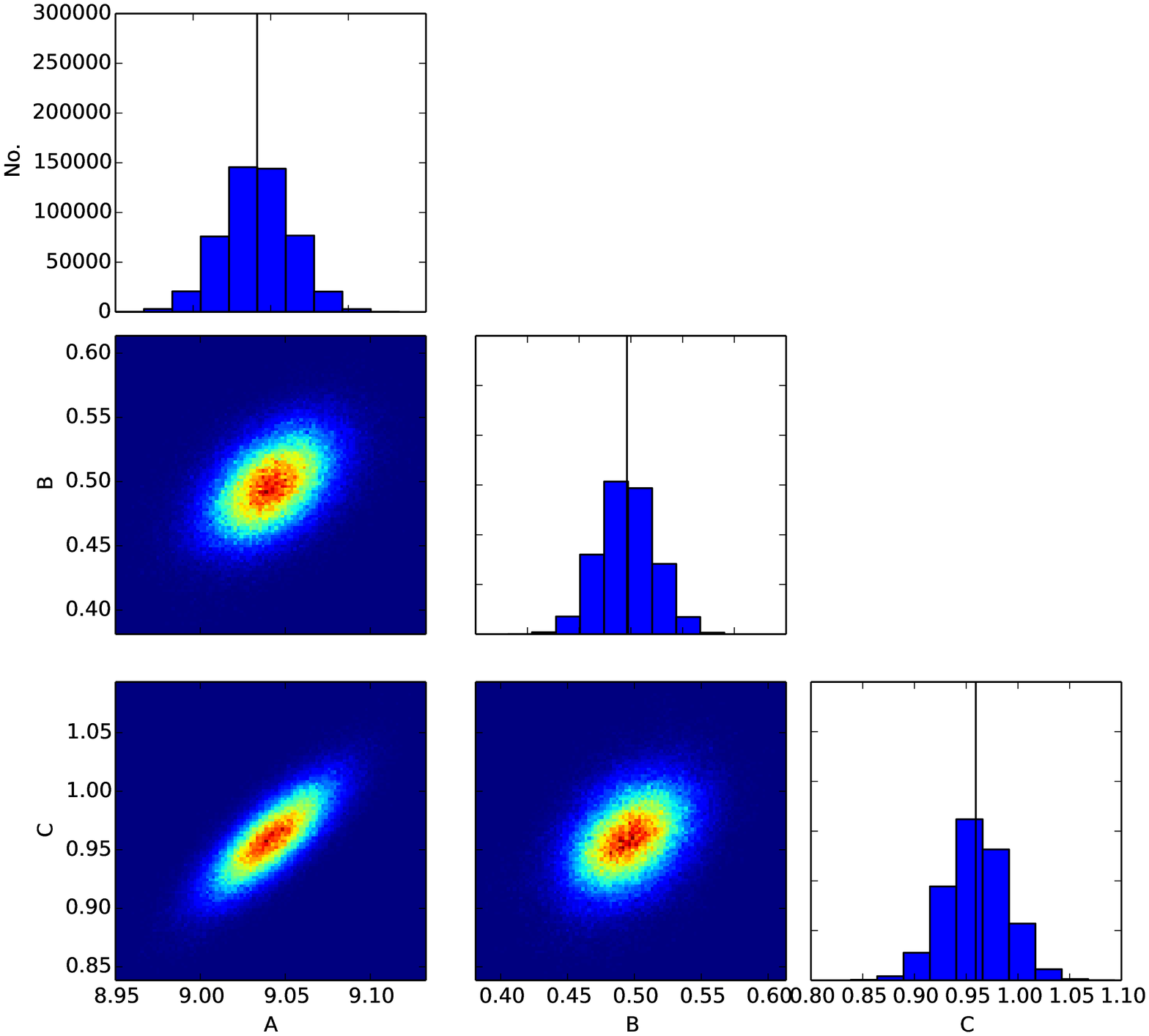}
  \caption{The results of fitting equation 4 to the data for the late-type
galaxies (as defined in Section 3).
The histograms show the probability distributions for the
three parameters, with the vertical lines showing the mean
values. The other panels show covariance plots.
}
\end{figure*}

%%%%%%%%%%%%%%%%%%%%%%%%%%%%%%%%%%%%%%%%%%%%%%%%%%

% Don't change these lines
\bsp	% typesetting comment
\label{lastpage}
\end{document}